\documentclass[3p,twocolumn]{elsarticle}

\usepackage{lineno,hyperref,amsmath,amssymb}
\modulolinenumbers[5]

\journal{Physics Letters B}

\usepackage{graphicx}
\usepackage{dcolumn}
\usepackage{bm}
\usepackage{hyperref}
\hypersetup{colorlinks=true, citecolor=blue, urlcolor=blue, linkcolor=blue}
\usepackage{color}
\usepackage{epsfig}
\usepackage{epstopdf}
\usepackage{amsmath}
\def\nuc#1#2{\relax\ifmmode{}^{#1}{\protect\text{#2}}\else${}^{#1}$#2\fi}
\definecolor{bobcatgreen}{rgb}{0.3,0.6,0.1}

\begin{document}
  \newcommand {\nc} {\newcommand}
  \nc {\beq} {\begin{eqnarray}}
  \nc {\eeq} {\nonumber \end{eqnarray}}
  \nc {\eeqn}[1] {\label {#1} \end{eqnarray}}
  \nc {\eol} {\nonumber \\}
  \nc {\eoln}[1] {\label {#1} \\}
  \nc {\ve} [1] {\mbox{\boldmath $#1$}}
  \nc {\ves} [1] {\mbox{\boldmath ${\scriptstyle #1}$}}
  \nc {\mrm} [1] {\mathrm{#1}}
  \nc {\half} {\mbox{$\frac{1}{2}$}}
  \nc {\thal} {\mbox{$\frac{3}{2}$}}
  \nc {\fial} {\mbox{$\frac{5}{2}$}}
  \nc {\la} {\mbox{$\langle$}}
  \nc {\ra} {\mbox{$\rangle$}}
  \nc {\etal} {\emph{et al.}}
  \nc {\eq} [1] {(\ref{#1})}
  \nc {\Eq} [1] {Eq.~(\ref{#1})}
  \nc {\Refc} [2] {Refs.~\cite[#1]{#2}}
  \nc {\Sec} [1] {Sec.~\ref{#1}}
  \nc {\chap} [1] {Chapter~\ref{#1}}
  \nc {\anx} [1] {Appendix~\ref{#1}}
  \nc {\tbl} [1] {Table~\ref{#1}}
  \nc {\Fig} [1] {Fig.~\ref{#1}}
  \nc {\ex} [1] {$^{#1}$}
  \nc {\Sch} {Schr\"odinger }
  \nc {\flim} [2] {\mathop{\longrightarrow}\limits_{{#1}\rightarrow{#2}}}
  \nc {\IR} [1]{\textcolor{red}{#1}}
  \nc {\IB} [1]{\textcolor{blue}{#1}}
  \nc{\IG}[1]{\textcolor{bobcatgreen}{#1}}
  \nc{\pderiv}[2]{\cfrac{\partial #1}{\partial #2}}
  \nc{\deriv}[2]{\cfrac{d#1}{d#2}}

\begin{frontmatter}

\title{Insensitivity of the Coulomb breakup of halo nuclei to spectroscopic factors}

\author[ou]{L.-P. Kubushishi\corref{cor1}}
\ead{lkubushi@ohio.edu}
\author[jgu]{P.~Capel}
\ead{pcapel@uni-mainz.de}

\cortext[cor1]{Corresponding author}
\address[ou]{Institute of Nuclear and Particle Physics and Department of Physics and Astronomy, Ohio University, Athens, OH 45701,USA}
\address[jgu]{Institut f\"ur Kernphysik, 
Johannes Gutenberg-Universit\"at Mainz,
Johann-Joachim-Becher Weg 45, 
D-55099 Mainz, Germany}

\begin{abstract}
Exotic nuclear structures such as halos are mostly studied using reactions.
In Coulomb breakup, the radioactive projectile dissociates through its interaction with a heavy target.
Often, a spectroscopic factor for the dominant core-halo configuration is inferred from experimental data.
In this work, we present a new calculation of the Coulomb breakup of the one-neutron halo nucleus $^{11}$Be performed with a coupled-channel effective particle-rotor model of that nucleus, which accounts for the excitation of the $^{10}$Be core.
Changes in the spectroscopic factor have no effect on the cross sections when the asymptotic normalisation coefficient is fixed, hence confirming the insensitivity of Coulomb-breakup cross sections to spectroscopic factors.
 \end{abstract}

\begin{keyword}
Halo nuclei; Coulomb breakup; spectroscopic factors; core excitation; effective coupled-channel model
\end{keyword}

\end{frontmatter}

\nolinenumbers

\section{Introduction}
In the 1980s, the development of radioactive-ion beams led to the discovery and study of halo nuclei \cite{Tan85b,Tan85l}.
These exotic nuclear structures exhibit a large matter radius compared to their isobars \cite{Tan96}.
This exceptional size results from the weak binding for one or two nucleons, which enables the valence nucleons to tunnel far from the core and form a diffuse halo around a compact core \cite{HJ87}.
Typical examples are $^{11}$Be and $^{15}$C for one-neutron halo nuclei, and $^{6}$He and $^{11}$Li for two-neutron halos.

Being located far from stability, halo nuclei are short lived, and their structure is mostly studied through reactions, such as breakup \cite{AN03,BC12}.
In breakup reactions, the halo dissociates from the core through the interaction with the target, hence unveiling the loosely-bound structure of the projectile.
From the analysis of data, a spectroscopic factor (SF) is usually inferred for the configuration, in which the halo neutron is bound in an $s$ or $p$ wave to the core in its ground state \cite{PAL03,FUK04}.
The SF is the square of the norm of the corresponding overlap wave function.

Although several articles have shown the non-observability of SFs in nuclear structure \cite{FH02,MKF15,MBF17}, this practice is widely used within the nuclear-reaction community.
Moreover, breakup reactions have been shown to be mostly peripheral in the sense that they probe mainly the tail of the projectile wave functions \cite{CAP06,CAP07,CAP18}.
Being quite insensitive to the internal part of the wave function, they cannot be used to infer its norm.
Instead, they are much better suited to deduce the asymptotic normalisation coefficient (ANC) of the core-halo wave function.
Since this conclusion has been drawn assuming overlap wave functions in a single core-halo channel with a unit norm, it 
remains to be confirmed assuming a more realistic coupled-channel description of the projectile.
Although some reaction models include such a description of halo nuclei \cite{SNT06,ML12}, this analysis has not yet been performed.

In the present work, we include the effective coupled-channel model of $^{11}$Be developed in Ref.~\cite{KC25b} in a first-order description of Coulomb breakup \cite{ALDWIN75,BERBAU88}.
This model accounts for the possible excitation of the $^{10}$Be towards its first excited $2^+$ state within a particle-rotor description of the nucleus \cite{ESB95,NUNES96,THOM04}.
By changing the coupling between the different core-neutron configurations, we vary their relative spectroscopic amplitudes, enabling us to assess their impact on the Coulomb-breakup cross section.
As an application, we consider the breakup of $^{11}$Be on $^{208}$Pb at 69~MeV/nucleon, which has been measured at RIKEN \cite{FUK04}.

After defining the effective particle-rotor model of $^{11}$Be, we briefly present how the Coulomb breakup is described at the first order of the perturbation, see \Sec{formalism}.
In \Sec{results} we present our results before drawing our conclusions in \Sec{conclusion}.

\section{Formalism}\label{formalism}
To compute the collision between $^{11}$Be and $^{208}$Pb, we consider the following three-body model.
The projectile $P$ is described as a neutron n loosely bound to a $^{10}$Be core $c$; it impinges on a $^{208}$Pb target $T$, of which the internal structure is neglected.

Following our previous work \cite{KC25b}, we improve the usual single-particle description of the projectile \cite{BC12} by including core excitation through a collective model \cite{BOHRMOTT69}.
The core is assumed to suffer a permanent quadrupole deformation and to be a rigid rotor \cite{ESB95,NUNES96,THOM04}.
The $c$-n Hamiltonian then reads 
\begin{equation}
 H(\ve{r},\xi)= -\frac{\hbar^2}{2\mu}\Delta + V_{c\rm n}({\ve{r}},\xi) + h_{c}({\xi}),
 \label{H_tot}
\end{equation}
where \ve{r} is the $c$-n relative coordinate, $\xi$ are the internal coordinates of the core, i.e., the Euler angles, $\mu = m_cm_{\rm n}/(m_c+m_{\rm n})$ is the $c$-n reduced mass, with $m_c$ and $m_{\rm n}$ the masses of the core and the neutron, respectively. 

The effective core-neutron potential $V_{c\rm n}$ is non-central, as it explicitly depends on the core degrees of freedom $\xi$.
As explained in Ref.~\cite{KC25b}, it is obtained by axially deforming the central $c$-n potential obtained within halo-effective field theory (Halo-EFT) at next-to-leading order \cite{CAP18}
\begin{equation}
    V(r;\sigma)=V_{J^\pi}^{(0)} e^{-\frac{r^2}{2\sigma^2}} + V_{J^\pi}^{(2)} r^2  e^{-\frac{r^2}{2\sigma^2}} \quad,
    \label{Vcn_NLO_central}
\end{equation}
where the range of the interaction $\sigma$ is an unfitted parameter corresponding to the short-distance physics neglected in this approach.
The low-energy constants (LECs) $V_{J^\pi}^{(0)}$ and $V_{J^\pi}^{(2)}$ are adjusted for each value of the total angular momentum $J$ and of the parity $\pi$ to known structure observables such as the binding energy and the ANC.
Invariably, effective-range parameters can also be used to constrain these LECs \cite{AP13}.

The non-central part of $V_{c\rm n}$, induced by the deformation of the core, translates into an angle dependence, which reads at first order
\begin{equation}
    V_{c\rm n}(\ve{r},\xi)= V(r;\sigma) + \beta\, \sigma\, Y_{2}^{0}(\hat{r}') \frac{d}{d\sigma}V(r;\sigma),
    \label{Vcpl_order1}
\end{equation}
where $\hat{r}'$ is the solid angle of the $c$-n relative coordinate $\ve{r}'$ expressed in the core's intrinsic reference frame, $Y_{\lambda}^{\mu}$ is a spherical harmonic and $\beta$ is the quadrupole deformation of the core, which acts here as a coupling strength between the different $c$-n configurations. In the derivative with respect to $\sigma$, we neglect the second term of $V$ \eq{Vcn_NLO_central}, as it yields a higher-order non-central correction to $V_{c\rm n}$.

In \Eq{H_tot}, the intrinsic Hamiltonian of the core $h_c$ accounts for the core structure in a collective model.
It satisfies the eigenvalue equation
\begin{equation}
   h_{c}({\xi})\phi_{M_{c}}^{I_{c}^{\pi_c}}(\xi)=\epsilon_{c}^{I_c^{\pi_c}} \phi_{M_{c}}^{I_{c}^{\pi_c}}(\xi) \quad,
   \label{H_core}
\end{equation}
where $\epsilon_c$ are the energies of the core states identified by its spin $I_c$, spin projection $M_c$ and parity $\pi_c$.
Since we describe the core as an axially symmetric rigid rotor, its eigenstates $\phi_{M_{c}}^{I_{c}\pi_c}$ are given by the Wigner matrices ${\cal D}^{I_c}_{M_cK_c}$ \cite{BOHRMOTT69}, where $K_c$ is the bandhead of the rotational band.

The eigenstates $\Psi^{J^\pi M}$ of the $c$-n Hamiltonian \eqref{H_tot} are expanded in a basis coupling the $c$-n relative motion to the internal structure of the core
\begin{equation}
    \Psi^{J^\pi M}(\ve{r},\xi)=\sum_{\alpha} i^\ell \frac{u_{\alpha}(r)}{r} [\mathcal{Y}_{\ell s j}(\hat{r}) \otimes \phi^{I_{c}\pi_c}(\xi)]^{J M},
    \label{totalwf}
\end{equation}
where $\alpha$=$\left\{n_r,\ell,s,j,I_c,\pi_c\right\}$ defines a channel, i.e., a set of quantum numbers describing each allowed $c$-n configuration.
In details, $n_r$ is the number of nodes in the radial wave function $u_\alpha$, $\ell$ is the orbital angular momentum of the $c$-n relative motion, $s=\half$ is the neutron spin and $j$ is the angular momentum resulting from the coupling of $\ell$ and $s$: $\ve{j} = \ve{\ell} + \ve{s}$.
This angular momentum $j$ is then coupled to the spin of the core $I_{c}$ to obtain the total angular momentum $J$ ($\ve{J} = \ve{j} + \ve{I_{c}}$).
The spin-angular part of the wave function $\mathcal{Y}_{lsjm}=[ Y_{\ell} \otimes \chi_s]^{j m}$ couples the spherical harmonic $Y_{\ell}^{m_\ell}$ to the spinor $\chi_s^{m_s}$.

Expanding the $c$-n Hamiltonian \eq{H_tot} in the set of basis functions \eq{totalwf} leads to coupled-channel equations for the reduced radial wave functions $u_{\alpha}$.
We solve these coupled equations for bound and scattering states using the R-matrix method on a Lagrange mesh \cite{DESC10,BAYE15}.
For bound states, the spectroscopic factor of a given channel $\alpha$ is defined as
\begin{equation}  {\cal S}_{\alpha}=\int_0^{\infty}\left|u{_\alpha}(r)\right|^2 dr \quad.
\label{defSF}
\end{equation}

In this exploratory step, we describe the Coulomb breakup of the projectile at the first order in perturbation theory \cite{ALDWIN75,BERBAU88}.
We assume the reaction to be dominated by a pure E1 transition, and we simulate the $P$-$T$ nuclear interaction by an impact-parameter cutoff $b_{\rm min}$.
The breakup cross section then reads
\begin{equation}\label{eq:dsigdE}
\frac{d\sigma_{\rm bu}}{dE}=\frac{16\pi^3}{9\hbar c}N_{{\rm E1}}(E,b_{\rm min})\frac{dB({\rm E1})}{dE} \quad,
\end{equation}
where $N_{{\rm E}1}$ is the number of equivalent photons exchanged between projectile and the target \cite{ALDWIN75,BERBAU88} and $dB({\rm E1})/dE$ is the reduced E1 transition probability computed between the projectile ground state and its continuum states.

Admittedly, higher-order effects are present \cite{EBS05,CB05} and the nuclear interaction affects the reaction process \cite{TS01,CBM03c}.
Nevertheless, the one-step E1 transition grasps most of the breakup strength, especially if the data are selected at forward angles \cite{CN17}, as will be done in the next section.

\begin{figure*}[ht]
    \centering
    \begin{minipage}{0.51\linewidth}
        \centering
        \includegraphics[width=\linewidth]
        {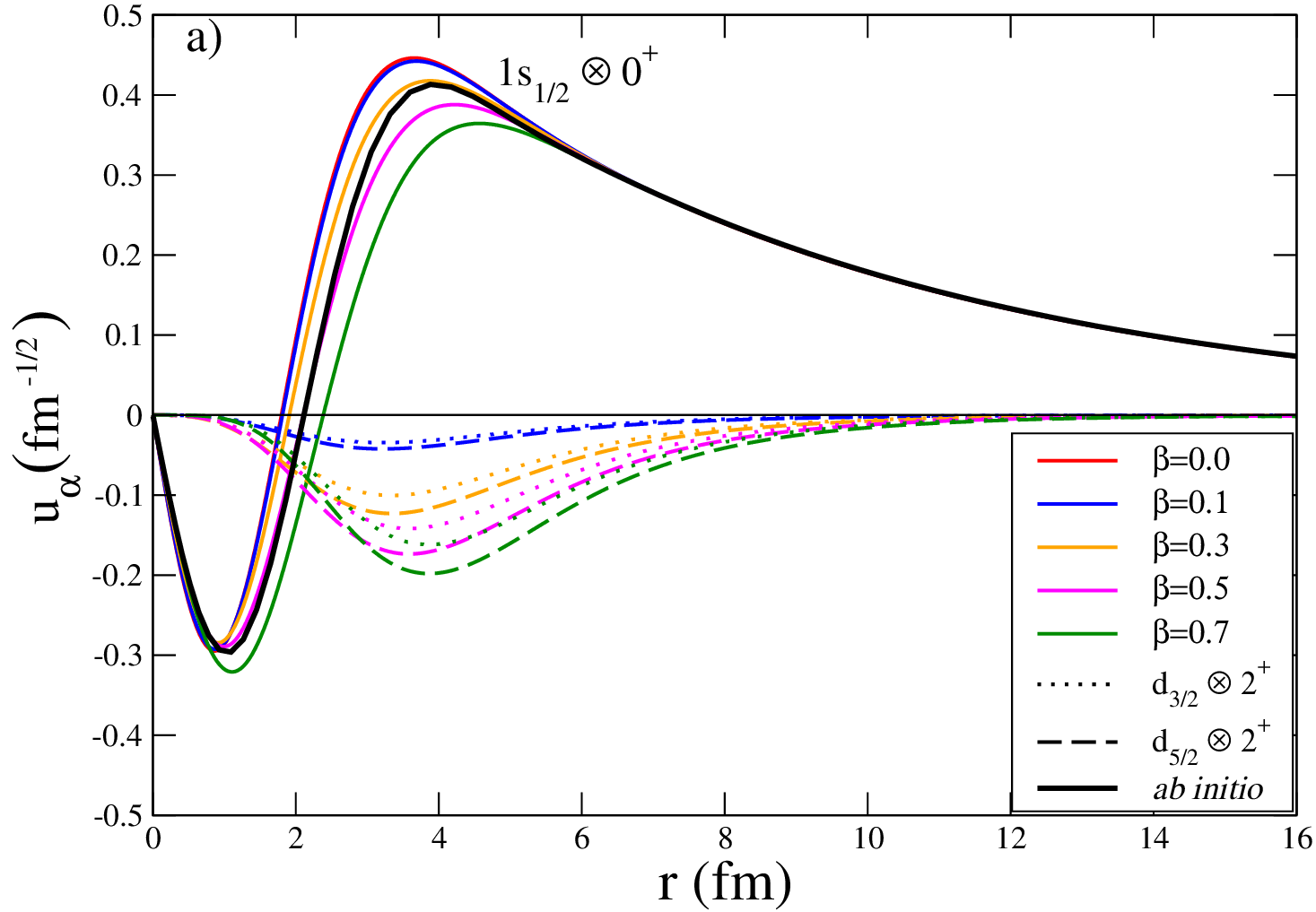}
    \end{minipage}\hfill
    \begin{minipage}{0.49\linewidth}
        \centering
        \includegraphics[width=\linewidth]{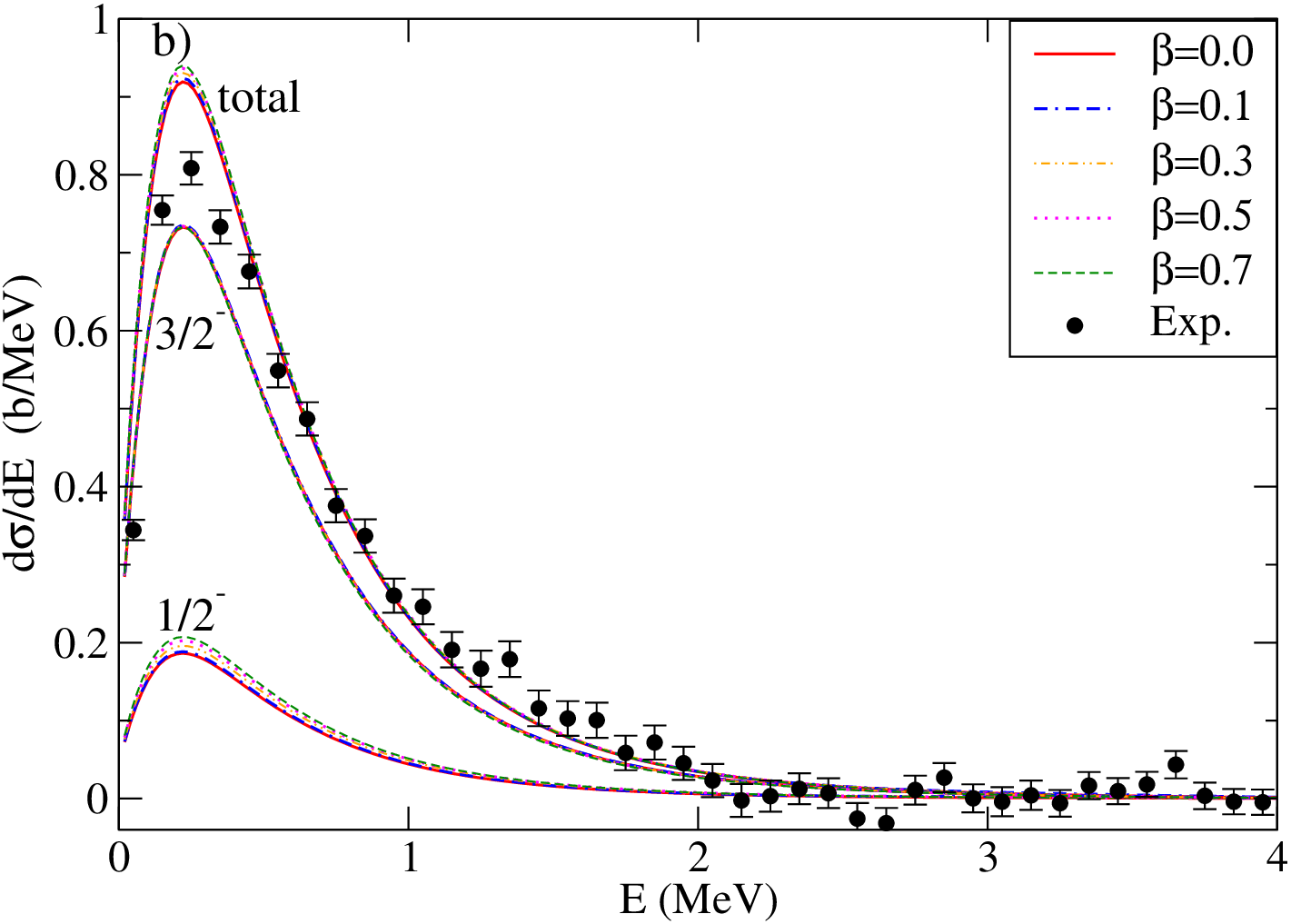}
    \end{minipage}
    \caption{Influence of the coupling between channels in the $^{11}$Be description.
(a) Radial wave functions of the $1s_{1/2} \otimes 0^+$ (solid lines), $0d_{5/2} \otimes 2^+$ (dashed lines), and $0d_{3/2} \otimes 2^+$ (dotted lines) configurations of the $\frac{1}{2}^+$ ground state of $^{11}$Be; the thick black line is the \textit{ab initio} prediction of Ref.~\cite{CAL16}.
(b) $d\sigma_{\rm bu}/dE$ for $^{11}$Be on Pb at 69~MeV/nucleon folded with the experimental resolution and compared to RIKEN data selected at forward angle \cite{FUK04}; the $\thal^-$ and $\half^-$ contributions are shown separately.
}
    \label{fig:plotab}
\end{figure*}

\begin{table*}[htb!]
\caption{\label{tabSFs}
Parameters of the effective $^{10}$Be-n potential \eqref{Vcn_NLO_central} in the
$\frac{1}{2}^+$ ground state for $\sigma=2$~fm.
For each value of $\beta$, we list the LECs reproducing the given $E_{1/2^+}$, ${\cal C}_{1/2^+}$, and ${\cal S}_{1s_{1/2}\otimes 0^+}$; \textit{ab initio} predictions are provided in the last line \cite{CAL16}.}
\centering
\begin{tabular}{cccccc}
\hline\hline
$\beta$ &
$V^{(0)}_{1/2^+}$ &
$V^{(2)}_{1/2^+}$ &
$E_{1/2^+}$ &
${\cal C}_{1/2^+}$ &
${\cal S}_{1s_{1/2}\otimes 0^+}$ \\
 & (MeV) & (MeV fm$^{-2}$) & (MeV) & (fm$^{-1/2}$) & \\ \hline
0   & $-80.755$  & $3.0$  & $-0.5031$ & $0.7845$ & 1.00 \\
0.1 & $-79.7078$ & $2.92$ & $-0.5031$ & $0.7861$ & 0.99 \\
0.3 & $-73.0924$ & $2.55$ & $-0.5031$ & $0.7859$ & 0.93 \\
0.5 & $-62.365$  & $2.01$ & $-0.5031$ & $0.7862$ & 0.85 \\
0.7 & $-51.2703$ & $1.45$ & $-0.5031$ & $0.7862$ & 0.81 \\ \hline
\textit{ab initio}& \cite{CAL16} &  & $-0.5$ & 0.786 & 0.90 \\
\hline\hline
\end{tabular}
\end{table*}

\section{Results}\label{results}
The $^{11}$Be spectrum exhibits a $J^{\pi}=\frac{1}{2}^+$ ground state bound by $S_{\rm n}(^{11}{\rm Be})=0.503$~MeV\footnote{The most recent value for that one-neutron separation energy is slightly different: $S_n(^{11}{\rm Be})=0.50164$~MeV$\pm0.25$~keV \cite{Masses21}. Nevertheless, we consider the same value as in Refs.~\cite{CAP18,KC25b} to ease the comparison with these previous calculations.} \cite{AJZEN90}.
The ANC of this state ${\cal C}_{1/2^+}=0.786$~fm$^{-1/2}$ is predicted \textit{ab initio} in Ref.~\cite{CAL16}.
This value of the ANC leads to excellent agreement with various reaction experiments \cite{CAP18,YC18,MC19,HC21,DEL23}.
As done in Refs.~\cite{ESB95,NUNES96}, we consider only the first two states of $^{10}$Be: its $0^+$ ground state and its first $2^+$ excited state at $\epsilon_{2^+}=3.368$~MeV \cite{AJZEN90}. 
We assume these states to be members of the same rotational band $K_c^{\pi_c}=0^+$, as confirmed by Refs.~\cite{KHD99,McCoy24}. 

Composing the angular momenta according to the coupling scheme in \Eq{totalwf}, the ground state $\frac{1}{2}^+$  of $^{11}$Be has three channels: $1s_{1/2} \otimes 0^+$, $d_{5/2} \otimes 2^+$ and $d_{3/2} \otimes 2^+$.
Following Ref.~\cite{CAP18}, we compute that ground-state wave function by fitting the LECs $V^{(0)}_{1/2^+}$ and $V^{(2)}_{1/2^+}$ to reproduce its experimental binding energy and \textit{ab initio} ANC.
We can change the coupling between the various channels by varying the quadrupole deformation $\beta$, see \Eq{Vcpl_order1}.
The values of these LECs, and the corresponding eigenenergie, ANC, and SF in the main channel ${\cal S}_{1s_{1/2}\otimes 0^+}$ are listed in Table~\ref{tabSFs} for values of $\beta$ between 0 (no coupling) and 0.7.
We choose the potential range $\sigma=2$~fm, because it leads to the largest admixture between configurations: up to 20\% at $\beta=0.7$.
Such a change in ${\cal S}_{1s_{1/2}\otimes 0^+}$ should be visible in breakup cross sections, were they sensitive to that value.

The radial wave functions in the three allowed channels of the $\frac{1}{2}^+$ ground state are shown in \Fig{fig:plotab}(a) for the different values of $\beta$.
The upper solid curves correspond to the main $1s_{1/2}\otimes0^+$ channel; they are compared to the \emph{ab initio} overlap wave function of Ref.~\cite{CAL16} (thick black line).
The lower curves correspond to the channels $0d_{5/2}\otimes2^+$ (dashed lines) and $0d_{3/2}\otimes2^+$ (dotted lines).

When $\beta$ increases, spectroscopic strength is transferred from the main channel to the $d$ waves coupled to the $2^+$ excited state of the $^{10}$Be core.
Because they dominate the long-range part of the wave function, and because the ANC is constrained by construction, the radial $1s_{1/2}\otimes0^+$ wave functions exhibit the same asymptotic behavior for all $\beta$.
The reduction in ${\cal S}_{1s_{1/2}\otimes 0^+}$ is therefore observed only for $r<6$\,fm.
It affects mainly the maximum of the wave function and is accompanied with a shift of the node towards larger radii.
The \emph{ab initio} overlap wave function is well reproduced by our model for $\beta\sim0.3$--0.5 \cite{KC25b}.

To compute the $dB({\rm E1})/dE$, we also need a description of the $^{10}$Be-n continuum.
As seen in Refs.~\cite{CAP06,SCB10}, the presence of the $\half^-$ excited bound state in the $^{11}$Be spectrum significantly constraints the corresponding continuum.
We therefore follow \cite{CAP18} and fit the LECs in the $\half^-$ wave to reproduce the experimental one-neutron separation energy and the \emph{ab initio} ANC of that state.
This leads to a fair agreement with the \emph{ab initio} $p_{1/2}$ phaseshift.
The \emph{ab initio} $p_{3/2}$ phaseshift being consistent with 0 at low energy \cite{CAL16}, we describe that continuum with plane waves as in Sec.~VI\,B of Ref.~\cite{CAP18}.
Tests performed with distorted $\thal^-$ waves as in Sec.~VIII\,A of Ref.~\cite{CAP18} show no significant difference with plane waves.
As already demonstrated with single-particle descriptions of $^{11}$Be \cite{CAP06}, the short-range part of the continuum wave functions has no significant influence of the computed cross section thanks to the peripherality of Coulomb breakup.

We then compute the differential Coulomb cross section $d\sigma_{\rm bu}/dE$ of $^{11}$Be on $^{208}$Pb at 69~MeV/nucleon using \Eq{eq:dsigdE} \cite{BERBAU88}.
We compare our results to the RIKEN data selected at forward angle ($\theta<1.3^\circ$) \cite{FUK04}.
Within a semiclassical approximation, this corresponds to setting an impact-parameter cutoff $b_{\rm min}\approx30$\,fm.
The corresponding cross sections, folded with the experimental resolution, are shown in Fig.~\ref{fig:plotab}(b) for different $\beta$s; the contributions of the $\thal^-$ and $\half^-$ partial waves are shown separately.
For all $\beta$, we obtain a good agreement with the experimental data, confirming the validity of our approach.
More importantly, the cross sections obtained with this coupled-channel description of $^{11}$Be are independent of the coupling, despite the 20\% change in ${\cal S}_{1s_{1/2}\otimes 0^+}$.
This independence is seen in both partial-wave contributions. We obtain the same results for other regulators $\sigma$, showing the generality of our results and their independence on the potential range.

\begin{figure*}[ht]
    \centering
    \begin{minipage}{0.51\linewidth}
        \centering
        \includegraphics[width=\linewidth]
        {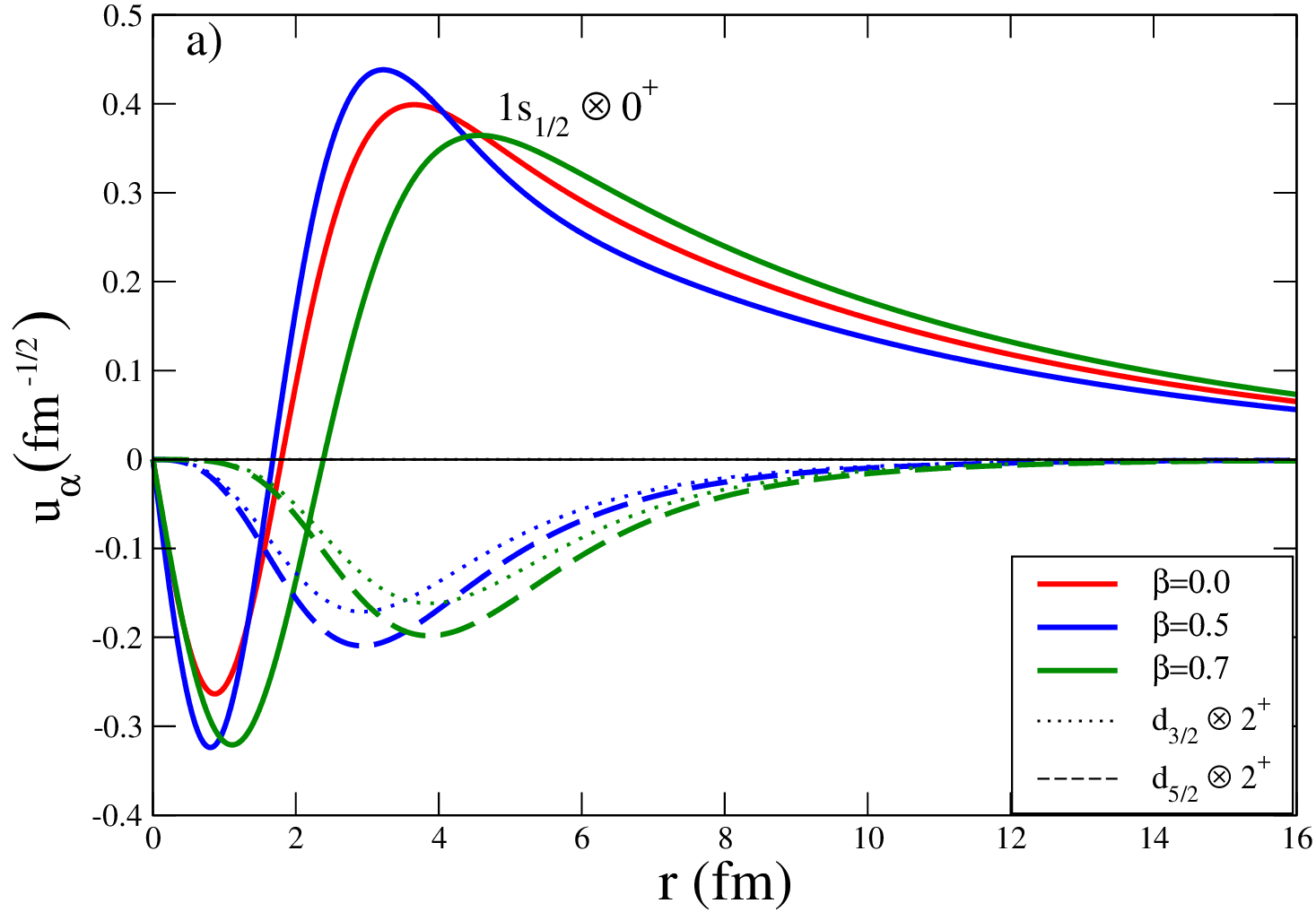}
    \end{minipage}\hfill
    \begin{minipage}{0.49\linewidth}
        \centering
        \includegraphics[width=\linewidth]{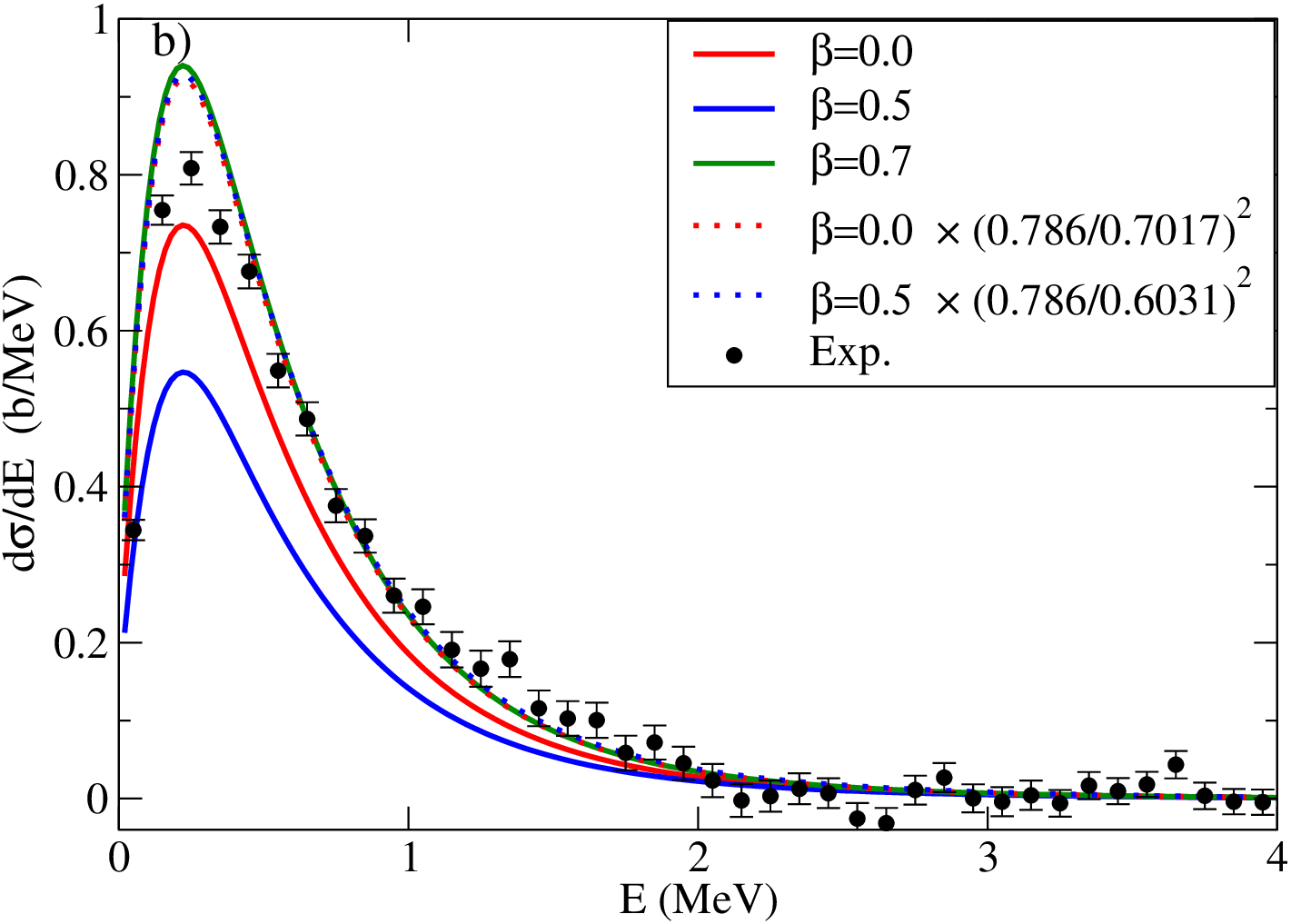}
    \end{minipage}
    \caption{Fixing ${\cal S}_{1s_{1/2}\otimes 0^+}=0.8$ for $\beta=0$ (red), 0.5 (blue), and 0.7 (green; same as in \Fig{fig:plotab}).
(a) $^{11}$Be ground-state wave functions in the $1s_{1/2} \otimes 0^+$ (solid lines), $0d_{5/2} \otimes 2^+$ (dashed lines), and $0d_{3/2} \otimes 2^+$ (dotted lines) channels.
(b) $d\sigma_{\rm bu}/dE$ for $^{11}$Be on Pb at 69~MeV/nucleon folded with the experimental resolution and compared to the RIKEN data \cite{FUK04}.
The dotted lines show the cross sections scaled to the \textit{ab initio} ${\cal C}_{1/2^+}=0.786$~fm$^{-1/2}$ \cite{CAL16} for $\beta=0$ and 0.5.
}
    \label{fig:plotcd}
\end{figure*}

\begin{table*}[htb!]
\caption{\label{tab2} Parameters of the effective $^{10}$Be-n potential \eqref{Vcn_NLO_central} in the $\frac{1}{2}^+$ ground state fitted to reproduce ${\cal S}_{1s_{1/2}\otimes 0^+}=0.8$ ($\sigma=2$\,fm).
For each value of $\beta$, we list the LECs and the resulting $E_{1/2^+}$, ${\cal S}_{1s_{1/2}\otimes 0^+}$, and ${\cal C}_{1/2^+}$.}
\centering
\begin{tabular}{cccccc}
\hline\hline
$\beta$ &
$V^{(0)}_{1/2^+}$ &
$V^{(2)}_{1/2^+}$ &
$E_{1/2^+}$ &
${\cal S}_{1s_{1/2}\otimes 0^+}$&
${\cal C}_{1/2^+}$\\
 & (MeV) & (MeV fm$^{-2}$) & (MeV) &  &(fm$^{-1/2}$) \\ \hline
0   & $-80.755$  & $3.0$  & $-0.5031$ & $0.80$ & $0.7017$\\
0.5 & $-93.984$  & $6.51$ & $-0.5031$ & $0.80$ & $0.6031$\\
0.7 & $-51.2703$ & $1.45$ & $-0.5031$ & $0.81$ & $0.7862$\\ \hline
\hline
\end{tabular}
\end{table*}

To further test the independence of the Coulomb breakup of $^{11}$Be of ${\cal S}_{1s_{1/2}\otimes 0^+}$, we perform another series of calculations.
This time, we fit the LECs in the $\half^+$ wave to $S_{\rm n}$ and ${\cal S}_{1s_{1/2}\otimes 0^+}=0.8$ (see Table~\ref{tab2}).
The latter value is consistent with the SF extracted from the RIKEN experiment \cite{FUK04}. 
Different $\beta$s then yield different ground-state wave functions.
In \Fig{fig:plotcd}(a), we plot these wave functions in the three channels: $1s_{1/2} \otimes 0^+$ (solid line), $0d_{5/2} \otimes 2^+$ (dashed lines), and $0d_{3/2} \otimes 2^+$ (dotted lines) obtained with $\beta=0$ (red), 0.5 (blue), and 0.7 (green).
The first one corresponds to a simple single-particle description of $^{11}$Be as used in Coulomb-breakup data analyses \cite{FUK04}.
Here it is merely normed to $\sqrt{0.8}$.
The third one is the same calculation as in \Fig{fig:plotab}.

Because the ANC is no longer constrained, we observe that each calculation leads to significantly different ${\cal C}_{1/2^+}$.
Since the norm of the $1s_{1/2} \otimes 0^+$ overlap wave function is fixed, larger ANCs naturally correspond to lower probability strength at small radii, and vice versa.

Were Coulomb-breakup cross sections sensitive to the SF of the dominant core-halo configuration, their value should be exactly the same for of all three descriptions of $^{11}$Be shown in \Fig{fig:plotcd}(a).
The results of our reaction calculations are plotted in \Fig{fig:plotcd}(b).
Despite having all been computed with ${\cal S}_{1s_{1/2}\otimes 0^+}=0.8$, the cross sections are by no means equal.
Rather they scale exactly with ${\cal C}_{1/2^+}^2$, as already predicted with single-particle descriptions of the projectile in Refs.~\cite{CAP06,CAP07,CAP18}, see the dotted lines in \Fig{fig:plotcd}(b).

\section{Conclusion}\label{conclusion}
Breakup reactions are used to probe the structure of halo nuclei \cite{PAL03,FUK04}.
Often a SF for the dominant core-halo configuration is inferred from measurements.
Previous theoretical studies have shown that these reactions are mostly peripheral in the sense that they probe mainly the tail of the projectile wave function, viz.\ the ANC and not the norm of the ground-state overlap wave function \cite{CAP06,CAP07,CAP18}.
Because these studies were performed with single-particle descriptions of the projectile, their conclusions need to be confirmed using a more realistic coupled-channel description.

In this Letter, we have presented a new calculation of the Coulomb breakup of $^{11}$Be using an effective  particle-rotor description of the one-neutron halo nucleus \cite{KC25b}.
We have investigated the influence of the coupling strength on the wave functions of the $\frac{1}{2}^+$ ground state of $^{11}$Be and computed the Coulomb breakup of $^{11}$Be on $^{208}$Pb at 69~MeV/nucleon at first order in perturbation theory.
The corresponding cross sections are in good agreement with existing data \cite{FUK04} once ${\cal C}_{1/2^+}$ and the phaseshift in the $c$-n continuum have been fitted to the \emph{ab initio} predictions of Ref.~\cite{CAL16}.
More importantly, it shows no dependence on ${\cal S}_{1s_{1/2}\otimes0^+}$.

We therefore confirm the conclusions drawn in Refs.~\cite{CAP06,CAP07,CAP18}: Coulomb breakup is insensitive to the SF of the dominant core-halo configuration.
In a near future, we plan to include the effective particle-rotor description of Ref.~\cite{KC25b} within a Coulomb-corrected eikonal reaction model \cite{CAPCCE08} to account for both higher-order effects and projectile-target nuclear interactions, which are missing in the present study.

\section*{Acknowledgments}
 This work was supported by the Deutsche Forschungsgemeinschaft (DFG, German Research Foundation) through Project-ID 279384907 – SFB 1245 and the Cluster of Excellence “Precision Physics, Fundamental Interactions and Structure of Matter” (PRISMA++ EXC 2118/2, Project ID No. 390831469), and by the U.S.~Department of Energy under contract No.~DE-FG02-93ER40756.

\bibliography{biblio}

@book{BOHRMOTT69,
  title     = "Nuclear Structure",
  author    = "A. Bohr and B. Mottelson",
  year      = 1969,
  publisher = "W. A. Benjamin",
  address   = "New York"
}

@misc{KC25b,
      title={Exploring core excitation in halo nuclei using halo effective field theory: an application to the bound states of $^{11}$Be}, 
      author={Live-Palm Kubushishi and Pierre Capel},
      year={2025},
      eprint={2507.13585},
      archivePrefix={arXiv},
      primaryClass={nucl-th},
      url={https://arxiv.org/abs/2507.13585}, 
}

@article{NUNES96,
journal = {Nucl. Phys. A},
volume = {596},
number = {2},
pages = {171-186},
year = {1996},
issn = {0375-9474},
doi = {https://doi.org/10.1016/0375-9474(95)00398-3},
url = {https://www.sciencedirect.com/science/article/pii/0375947495003983},
author = {F.M. Nunes and I.J. Thompson and R.C. Johnson},
title = {Core excitation in one neutron halo systems}
}

@article{THOM04,
title = {{FaCE}: a tool for three body {F}addeev calculations with core excitation},
author = {I.J. Thompson and F.M. Nunes and B.V. Danilin},
journal = {Comput. Phys. Comm.},
volume = {161},
number = {1},
pages = {87-107},
year = {2004},
issn = {0010-4655},
doi = {https://doi.org/10.1016/j.cpc.2004.03.007},
url = {https://www.sciencedirect.com/science/article/pii/S0010465504002140}
}

@article{BERBAU88,
title = {Electromagnetic processes in relativistic heavy ion collisions},
journal = {Physics Reports},
volume = {163},
number = {5},
pages = {299-408},
year = {1988},
issn = {0370-1573},
doi = {https://doi.org/10.1016/0370-1573(88)90142-1},
url = {https://www.sciencedirect.com/science/article/pii/0370157388901421},
author = {Carlos A. Bertulani and Gerhard Baur}
}

@book{ALDWIN75,
  title={Electromagnetic Excitation: Theory of Coulomb Excitation with Heavy Ions},
  author={Alder, K. and Winther, A.},
  isbn={9780444108265},
  lccn={73091445},
  url={https://books.google.com/books?id=azet0AEACAAJ},
  year={1975},
  publisher={North-Holland Publishing Company}
}

@article{CAP18,
  title = {Dissecting reaction calculations using halo effective field theory and ab initio input},
  author = {Capel, P. and Phillips, D. R. and Hammer, H.-W.},
  journal = {Phys. Rev. C},
  volume = {98},
  issue = {3},
  pages = {034610},
  numpages = {17},
  year = {2018},
  month = {Sep},
  publisher = {American Physical Society},
  doi = {10.1103/PhysRevC.98.034610},
  url = {https://link.aps.org/doi/10.1103/PhysRevC.98.034610}
}

@article{DESC10,
doi = {10.1088/0034-4885/73/3/036301},
url = {https://dx.doi.org/10.1088/0034-4885/73/3/036301},
year = {2010},
month = {feb},
publisher = {},
volume = {73},
number = {3},
pages = {036301},
author = {P Descouvemont and D Baye},
title = {{The R-matrix theory}},
journal = {Reports on Progress in Physics}
}

@article{BAYE15,
title = {{The Lagrange-mesh method}},
journal = {Physics Reports},
volume = {565},
pages = {1-107},
year = {2015},
note = {The Lagrange-mesh method},
issn = {0370-1573},
doi = {https://doi.org/10.1016/j.physrep.2014.11.006},
url = {https://www.sciencedirect.com/science/article/pii/S0370157314004086},
author = {Daniel Baye}
}

@article{PAL03,
  title = {Exclusive measurement of breakup reactions with the one-neutron halo nucleus ${}^{11}\mathrm{Be}$},
  author = {Palit, R. and Adrich, P. and Aumann, T. and Boretzky, K. and Carlson, B. V. and Cortina, D. and Datta Pramanik, U. and Elze, Th. W. and Emling, H. and Geissel, H. and Hellstr\"om, M. and Jones, K. L. and Kratz, J. V. and Kulessa, R. and Leifels, Y. and Leistenschneider, A. and M\"unzenberg, G. and Nociforo, C. and Reiter, P. and Simon, H. and S\"ummerer, K. and Walus, W.},
  collaboration = {LAND/FRS Collaboration},
  journal = {Phys. Rev. C},
  volume = {68},
  issue = {3},
  pages = {034318},
  numpages = {14},
  year = {2003},
  month = {Sep},
  publisher = {American Physical Society},
  doi = {10.1103/PhysRevC.68.034318},
  url = {https://link.aps.org/doi/10.1103/PhysRevC.68.034318}
}

@article{FUK04,
  title = {Coulomb and nuclear breakup of a halo nucleus $^{11}\mathrm{Be}$},
  author = {Fukuda, N. and Nakamura, T. and Aoi, N. and Imai, N. and Ishihara, M. and Kobayashi, T. and Iwasaki, H. and Kubo, T. and Mengoni, A. and Notani, M. and Otsu, H. and Sakurai, H. and Shimoura, S. and Teranishi, T. and Watanabe, Y. X. and Yoneda, K.},
  journal = {Phys. Rev. C},
  volume = {70},
  issue = {5},
  pages = {054606},
  numpages = {12},
  year = {2004},
  month = {Nov},
  publisher = {American Physical Society},
  doi = {10.1103/PhysRevC.70.054606},
  url = {https://link.aps.org/doi/10.1103/PhysRevC.70.054606}
}

@conference{McCoy24,
      title={The beryllium isotopic chain: evolution of structure in neutron rich nuclei}, 
      booktitle={HaloWeek'24 - nuclei at and beyond the driplines},
      author={Anna McCoy},
      address={Chalmers University of Technology, Gothenburg, Sweden},
      year={2024},
      url={https://indico.cern.ch/event/1319370/contributions/5984436/}, 
}

@article{KHD99,
  title = {Structure of excited states of ${}^{10}\mathrm{Be}$ studied with antisymmetrized molecular dynamics},
  author = {Kanada-En'yo, Y. and Horiuchi, H. and Dot\'e, A.},
  journal = {Phys. Rev. C},
  volume = {60},
  issue = {6},
  pages = {064304},
  numpages = {10},
  year = {1999},
  month = {Oct},
  publisher = {American Physical Society},
  doi = {10.1103/PhysRevC.60.064304},
  url = {https://link.aps.org/doi/10.1103/PhysRevC.60.064304}
}

@article{CAL16,
  title = {Can Ab Initio Theory Explain the Phenomenon of Parity Inversion in $^{11}\mathrm{Be}$?},
  author = {Calci, Angelo and Navr\'atil, Petr and Roth, Robert and Dohet-Eraly, J\'er\'emy and Quaglioni, Sofia and Hupin, Guillaume},
  journal = {Phys. Rev. Lett.},
  volume = {117},
  issue = {24},
  pages = {242501},
  numpages = {6},
  year = {2016},
  month = {Dec},
  publisher = {American Physical Society},
  doi = {10.1103/PhysRevLett.117.242501},
  url = {https://link.aps.org/doi/10.1103/PhysRevLett.117.242501}
}

@article{SNT06,
  title = {Core transitions in the breakup of exotic nuclei},
  author = {Summers, N. C. and Nunes, F. M. and Thompson, I. J.},
  journal = {Phys. Rev. C},
  volume = {73},
  issue = {3},
  pages = {031603},
  numpages = {4},
  year = {2006},
  month = {Mar},
  publisher = {American Physical Society},
  doi = {10.1103/PhysRevC.73.031603},
  url = {https://link.aps.org/doi/10.1103/PhysRevC.73.031603}
}

@article{ML12,
  title = {Interplay Between Valence and Core Excitation Mechanisms in the Breakup of Halo Nuclei},
  author = {Moro, A. M. and Lay, J. A.},
  journal = {Phys. Rev. Lett.},
  volume = {109},
  issue = {23},
  pages = {232502},
  numpages = {5},
  year = {2012},
  publisher = {American Physical Society},
  doi = {10.1103/PhysRevLett.109.232502},
  url = {https://link.aps.org/doi/10.1103/PhysRevLett.109.232502}
}

@article{Masses21,
	doi = {10.1088/1674-1137/abddae},
	url = {https://doi.org/10.1088/1674-1137/abddae},
	year = 2021,
	month = {mar},
	publisher = {{IOP} Publishing},
	volume = {45},
	number = {3},
	pages = {030001},
	author = {F.G. Kondev and M. Wang and W.J. Huang and S. Naimi and G. Audi},
	title = {The {NUBASE}2020 evaluation of nuclear physics properties},
	journal = {Chinese Phys. C}
}

@article{AJZEN90,
title = {{Energy levels of light nuclei A = 11--12}},
journal = {Nuclear Physics A},
volume = {506},
number = {1},
pages = {1-158},
year = {1990},
issn = {0375-9474},
doi = {https://doi.org/10.1016/0375-9474(90)90271-M},
url = {https://www.sciencedirect.com/science/article/pii/037594749090271M},
author = {F. Ajzenberg-Selove}
}

@article{MC19,
title = {Reliable extraction of the $dB({\rm E1})/dE$ for $^{11}${Be} from its breakup at 520 MeV/nucleon},
journal = {Phys. Lett. B},
volume = {790},
pages = {367-371},
year = {2019},
issn = {0370-2693},
doi = {https://doi.org/10.1016/j.physletb.2019.01.041},
url = {https://www.sciencedirect.com/science/article/pii/S0370269319300577},
author = {L. Moschini and P. Capel}
}

@article{YC18,
  title = {Systematic analysis of the peripherality of the $^{10}\mathrm{Be}(d,p)^{11}\mathrm{Be}$ transfer reaction and extraction of the asymptotic normalization coefficient of $^{11}\mathrm{Be}$ bound states},
  author = {Yang, J. and Capel, P.},
  journal = {Phys. Rev. C},
  volume = {98},
  issue = {5},
  pages = {054602},
  numpages = {10},
  year = {2018},
  month = {Nov},
  publisher = {American Physical Society},
  doi = {10.1103/PhysRevC.98.054602},
  url = {https://link.aps.org/doi/10.1103/PhysRevC.98.054602}
}

@article{HC21,
  title = {Halo effective field theory analysis of one-neutron knockout reactions of $^{11}\mathrm{Be}$ and $^{15}\mathrm{C}$},
  author = {Hebborn, C. and Capel, P.},
  journal = {Phys. Rev. C},
  volume = {104},
  issue = {2},
  pages = {024616},
  numpages = {7},
  year = {2021},
  month = {Aug},
  publisher = {American Physical Society},
  doi = {10.1103/PhysRevC.104.024616},
  url = {https://link.aps.org/doi/10.1103/PhysRevC.104.024616}
}

@article{DEL23,
title = {Nonlocal optical potential with core excitation in Be10(d,p)11Be and Be11(p,d)10Be reactions},
journal = {Physics Letters B},
volume = {840},
pages = {137867},
year = {2023},
issn = {0370-2693},
doi = {https://doi.org/10.1016/j.physletb.2023.137867},
url = {https://www.sciencedirect.com/science/article/pii/S0370269323002010},
author = {A. Deltuva and D. Jurčiukonis}
}

@article{ESB95,
  title = {Positive parity states in $^{11}\mathrm{Be}$},
  author = {Esbensen, H. and Brown, B. A. and Sagawa, H.},
  journal = {Phys. Rev. C},
  volume = {51},
  issue = {3},
  pages = {1274--1279},
  numpages = {0},
  year = {1995},
  month = {Mar},
  publisher = {American Physical Society},
  doi = {10.1103/PhysRevC.51.1274},
  url = {https://link.aps.org/doi/10.1103/PhysRevC.51.1274}
}

@article{CAP06,
  title = {Influence of the projectile description on breakup calculations},
  author = {Capel, P. and Nunes, F. M.},
  journal = {Phys. Rev. C},
  volume = {73},
  issue = {1},
  pages = {014615},
  numpages = {9},
  year = {2006},
  month = {Jan},
  publisher = {American Physical Society},
  doi = {10.1103/PhysRevC.73.014615},
  url = {https://link.aps.org/doi/10.1103/PhysRevC.73.014615}
}

@article{CAP07,
  title = {Peripherality of breakup reactions},
  author = {Capel, P. and Nunes, F. M.},
  journal = {Phys. Rev. C},
  volume = {75},
  issue = {5},
  pages = {054609},
  numpages = {6},
  year = {2007},
  month = {May},
  publisher = {American Physical Society},
  doi = {10.1103/PhysRevC.75.054609},
  url = {https://link.aps.org/doi/10.1103/PhysRevC.75.054609}
}

@article{CAPCCE08,
  title = {Coulomb-corrected eikonal description of the breakup of halo nuclei},
  author = {Capel, P. and Baye, D. and Suzuki, Y.},
  journal = {Phys. Rev. C},
  volume = {78},
  issue = {5},
  pages = {054602},
  numpages = {10},
  year = {2008},
  month = {Nov},
  publisher = {American Physical Society},
  doi = {10.1103/PhysRevC.78.054602},
  url = {https://link.aps.org/doi/10.1103/PhysRevC.78.054602}
}

@article{SCB10,
  title = {Influence of low-energy scattering on loosely bound states},
  author = {Sparenberg, Jean-Marc and Capel, Pierre and Baye, Daniel},
  journal = {Phys. Rev. C},
  volume = {81},
  issue = {1},
  pages = {011601},
  numpages = {4},
  year = {2010},
  month = {Jan},
  publisher = {American Physical Society},
  doi = {10.1103/PhysRevC.81.011601},
  url = {https://link.aps.org/doi/10.1103/PhysRevC.81.011601}
}

@article{Tan96,
doi = {10.1088/0954-3899/22/2/004},
url = {https://dx.doi.org/10.1088/0954-3899/22/2/004},
year = {1996},
month = {feb},
publisher = {},
volume = {22},
number = {2},
pages = {157},
author = {Isao Tanihata},
title = {Neutron halo nuclei},
journal = {J. Phys. G}
}

@article{Tan85b,
title = {Measurements of interaction cross sections and radii of He isotopes},
journal = {Phys. Lett.},
volume = {B160},
number = {6},
pages = {380-384},
year = {1985},
issn = {0370-2693},
doi = {https://doi.org/10.1016/0370-2693(85)90005-X},
url = {https://www.sciencedirect.com/science/article/pii/037026938590005X},
author = {I. Tanihata and H. Hamagaki and O. Hashimoto and S. Nagamiya and Y. Shida and N. Yoshikawa and O. Yamakawa and K. Sugimoto and T. Kobayashi and D.E. Greiner and N. Takahashi and Y. Nojiri}
}

@article{Tan85l,
  title = {Measurements of Interaction Cross Sections and Nuclear Radii in the Light $p$-Shell Region},
  author = {Tanihata, I. and Hamagaki, H. and Hashimoto, O. and Shida, Y. and Yoshikawa, N. and Sugimoto, K. and Yamakawa, O. and Kobayashi, T. and Takahashi, N.},
  journal = {Phys. Rev. Lett.},
  volume = {55},
  issue = {24},
  pages = {2676--2679},
  numpages = {0},
  year = {1985},
  month = {Dec},
  publisher = {American Physical Society},
  doi = {10.1103/PhysRevLett.55.2676},
  url = {https://link.aps.org/doi/10.1103/PhysRevLett.55.2676}
}

@article{HJ87,
doi = {10.1209/0295-5075/4/4/005},
url = {https://dx.doi.org/10.1209/0295-5075/4/4/005},
year = {1987},
month = {aug},
publisher = {},
volume = {4},
number = {4},
pages = {409},
author = {P. G. Hansen and  B. Jonson},
title = {The Neutron Halo of Extremely Neutron-Rich Nuclei},
journal = {Europhys. Lett.}
}

@article{AN03,
doi = {10.1088/0954-3899/29/11/R01},
url = {https://dx.doi.org/10.1088/0954-3899/29/11/R01},
year = {2003},
month = {oct},
publisher = {},
volume = {29},
number = {11},
pages = {R89},
author = {Jim Al-Khalili and  Filomena Nunes},
title = {Reaction models to probe the structure of light exotic nuclei},
journal = {J. Phys. G}
}

@article{BC12,
  title = {Breakup reaction models for two- and three-cluster projectiles},
  author = {D. Baye and P. Capel},
  note = {{Ed. C. Beck}},
  booktitle = {Clusters in Nuclei, Vol. 2},
  journal = {Lecture Notes in Physics},
  volume = {848},
  pages = {121--163},
  publisher = {Springer},
  address = {Heidelberg},
  year = {2012},
  doi = {10.1007/978-3-642-24707-1_3},
  url = {https://doi.org/10.1007/978-3-642-24707-1_3}
}

@article{TS01,
  title = {Dynamical description of the breakup of one-neutron halo nuclei ${}^{11}\mathrm{Be}$ and ${}^{19}\mathrm{C}$},
  author = {Typel, S. and Shyam, R.},
  journal = {Phys. Rev. C},
  volume = {64},
  issue = {2},
  pages = {024605},
  numpages = {8},
  year = {2001},
  month = {Jun},
  publisher = {American Physical Society},
  doi = {10.1103/PhysRevC.64.024605},
  url = {https://link.aps.org/doi/10.1103/PhysRevC.64.024605}
}

@article{CBM03c,
  title = {Time-dependent analysis of the breakup of halo nuclei},
  author = {Capel, P. and Baye, D. and Melezhik, V. S.},
  journal = {Phys. Rev. C},
  volume = {68},
  issue = {1},
  pages = {014612},
  numpages = {13},
  year = {2003},
  month = {Jul},
  publisher = {American Physical Society},
  doi = {10.1103/PhysRevC.68.014612},
  url = {https://link.aps.org/doi/10.1103/PhysRevC.68.014612}
}

@article{CB05,
  title = {{Coupling-in-the-continuum effects in Coulomb dissociation of halo nuclei}},
  author = {Capel, P. and Baye, D.},
  journal = {Phys. Rev. C},
  volume = {71},
  issue = {4},
  pages = {044609},
  numpages = {7},
  year = {2005},
  month = {Apr},
  publisher = {American Physical Society},
  doi = {10.1103/PhysRevC.71.044609},
  url = {https://link.aps.org/doi/10.1103/PhysRevC.71.044609}
}

@article{EBS05,
  title = {Reconciling Coulomb Dissociation and Radiative Capture Measurements},
  author = {Esbensen, H. and Bertsch, G. F. and Snover, K. A.},
  journal = {Phys. Rev. Lett.},
  volume = {94},
  issue = {4},
  pages = {042502},
  numpages = {4},
  year = {2005},
  month = {Jan},
  publisher = {American Physical Society},
  doi = {10.1103/PhysRevLett.94.042502},
  url = {https://link.aps.org/doi/10.1103/PhysRevLett.94.042502}
}

@article{CN17,
  title = {{Reconciling Coulomb breakup and neutron radiative capture}},
  author = {Capel, P. and Nollet, Y.},
  journal = {Phys. Rev. C},
  volume = {96},
  issue = {1},
  pages = {015801},
  numpages = {8},
  year = {2017},
  month = {Jul},
  publisher = {American Physical Society},
  doi = {10.1103/PhysRevC.96.015801},
  url = {https://link.aps.org/doi/10.1103/PhysRevC.96.015801}
}

@article{FH02,
title = {Are occupation numbers observable?},
journal = {Phys. Lett. B},
volume = {531},
number = {3},
pages = {203-208},
year = {2002},
issn = {0370-2693},
doi = {https://doi.org/10.1016/S0370-2693(01)01504-0},
url = {https://www.sciencedirect.com/science/article/pii/S0370269301015040},
author = {R.J. Furnstahl and H.-W. Hammer}
}

@article{MKF15,
  title = {Deuteron electrodisintegration with unitarily evolved potentials},
  author = {More, S. N. and K\"onig, S. and Furnstahl, R. J. and Hebeler, K.},
  journal = {Phys. Rev. C},
  volume = {92},
  issue = {6},
  pages = {064002},
  numpages = {16},
  year = {2015},
  month = {Dec},
  publisher = {American Physical Society},
  doi = {10.1103/PhysRevC.92.064002},
  url = {https://link.aps.org/doi/10.1103/PhysRevC.92.064002}
}

@article{MBF17,
  title = {Scale dependence of deuteron electrodisintegration},
  author = {More, S. N. and Bogner, S. K. and Furnstahl, R. J.},
  journal = {Phys. Rev. C},
  volume = {96},
  issue = {5},
  pages = {054004},
  numpages = {15},
  year = {2017},
  month = {Nov},
  publisher = {American Physical Society},
  doi = {10.1103/PhysRevC.96.054004},
  url = {https://link.aps.org/doi/10.1103/PhysRevC.96.054004}
}

@article{AP13,
title = {19C in halo EFT: Effective-range parameters from Coulomb dissociation experiments},
journal = {Nucl. Phys. A},
volume = {913},
pages = {103-115},
year = {2013},
issn = {0375-9474},
doi = {https://doi.org/10.1016/j.nuclphysa.2013.05.021},
url = {https://www.sciencedirect.com/science/article/pii/S0375947413006040},
author = {B. Acharya and Daniel R. Phillips}
}
\end{document}